\documentclass[12pt]{article}

\date{}

\usepackage{amsmath}
\usepackage{graphicx}

\title{ Deconfinement  transition for  nonzero baryon density in the  Field Correlator  Method.}
\author{ Yu.A.Simonov, M.A.Trusov\\
State Research
Center\\Institute of Theoretical and Experimental Physics, \\
Moscow, 117218 Russia}

\newcommand{\beq}{\begin{eqnarray}}
 \newcommand{\eeq}{\end{eqnarray}}
\newcommand{\be}{\begin{equation}}
 \newcommand{\ee}{\end{equation}}

 \def\la{\mathrel{\mathpalette\fun <}}

\def\fun#1#2{\lower3.6pt\vbox{\baselineskip0pt\lineskip.9pt
\ialign{$\mathsurround=0pt#1\hfil ##\hfil$\crcr#2\crcr\sim\crcr}}}

\newcommand{{\SD}}{\rm SD}

\newcommand{\lan}{\langle}
\newcommand{\ran}{\rangle}

\begin{document}
\maketitle
\begin{abstract}

Deconfinement phase transition due to disappearance of confining
colorelectric field correlators is described using nonperturbative
equation of state. The resulting transition temperature $T_c(\mu)$
at any chemical potential $\mu$ is  expressed in terms of the
change of gluonic condensate $\Delta G_2$ and absolute value of
Polyakov loop $L_{fund} (T_c)$, known from lattice and analytic
data, and is in good agreement with lattice data for $\Delta G_2
\approx 0.0035 $ GeV$^4$. E.g. $T_c(0) =0.27; 0.19; 0.17$ GeV for
$n_f=0,2,3$ respectively.

\end{abstract}

\section{Introduction}

1. Phase transition at nonzero $\mu$ and dynamics of quark gluon
plasma (qgp) is now of great interest because of impressive
results of heavy ion  experiments, see \cite{1} for a recent
review and references. The topic calls for a nonperturbative (NP)
treatment of QCD degrees of freedom at nonzero $T$ and $\mu$,
which is especially important for not very large $T,\mu$.  Below
we are using the NP approach based on  Field Correlator Method
(FCM) \cite{2}, which was applied to nonzero $T$ in \cite{3,4}.

The main advantage of FCM is a natural explanation and treatment
of dynamics of confinement, as well as the deconfinement
transition \cite{3,4}, in terms of Color Electric (CE) $D^E(x),
D^E_1(x)$  and Color Magnetic (CM) Gaussian (quadratic in
$F_{\mu\nu}^a$) correlators $D^H(x), D_1^H(x)$.

The correlators $D^E(x)$ and $D^H(x)$ ensure confinement in the
planes $(4,i)$ and $(i,k)$ respectively, $i,k=1,2,3$ so that
standard string tension $\sigma^E\equiv \sigma =\frac12 \int
D^E(x) d^2 x$, and spatial string tension $\sigma^H\equiv
\sigma_s=\frac12 \int D^H(x) d^2x$. Correlators $D^E_1, D^H_1$
contain perturbative series, and $D_1^E$ plays an important role
in that it contributes to the  modulus of the  Polyakov line; at
$T\geq T_c$ one has \cite{5}

\be L_{fund}^{(V)} =|\frac{1}{N_c} \lan tr P \exp  ig \int^\beta_0
A_4 dz_4\ran | = \exp \left(-\frac{V_1
(\infty,T)}{2T}\right)\label{1}\ee with the static $Q\bar Q$
potential

\be V_1 (r,T) = \int^\beta_0 d\nu ( 1-\nu T) \int^r_0  \xi  d \xi
D^E_1 (\sqrt{\xi^2 +\nu^2}).\label{2}\ee

Deconfinement phase transition in this language  is the
disappearance of $D^E(x)$ (and $\sigma^E$) at $T\geq T_c$, while
$L_{fund}^{(V)}$ and $L_{adj}^{(V)}= (L_{fund}^{(V)})^{9/4}$ are
nonzero there\footnote{ The subscript $V$ in $L_{fund}^{(V)}$ is
to distinguish from  $L_{fund}^{(F)} $ calculated on the lattice
with singlet free energy $F^1_{Q\bar Q} (\infty, T)$ replacing
$V_1(\infty, T)$ in (\ref{1}). It is clear that $L^{(F)} $
contains all bound states in addition to the ground state
(selfenergy) $V_1(\infty, T)$, hence $F^1_{Q\bar Q} \leq V_1$. We
shall ignore the difference in the first approximation  in  what
follows and write $L_{fund}$.}.

However the disappearance  of  $D^E(x) =\frac{1}{N_c} \lan tr E_i
(x) \Phi (x,y) E_{i}(y)\ran$ implies vanishing of a part of vacuum
energy density, \be \varepsilon_{vac} = \frac{\beta(\alpha_s)}{16
\alpha_s}\lan (F^a_{\mu\nu})^2\ran =- \frac{(11-\frac23 n_f)}{32}
G_2,~~ D^E(0) + D_1^E(0) + D^H(0) + D_1^H(0) =\frac{\pi^2}{9}
G_2\label{3}\ee and $G_2$ is the gluon condensate \cite{6}. At
$T=0$, $D^E=D^H, D^E_1 =D^H_1$ and for $T>0$ both $ D^H, D_1^H$ do
not change till $T\approx 2 T_c$, while $D^E$ disappears at $T\geq
T_c$ \cite{7} in agreement with the deconfinement  mechanism
suggested in \cite{4}.

Particle data \cite{7} and analytic study \cite{8} imply that
$D_1^{(E,H)} (x) \approx 0.2 D^{(E,H)} (x)$, therefore one expects
that $\Delta G_2 = G_2 (T<T_c)-G_2 (T>T_c)\approx \frac12
G_2(T<T_c) \approx \frac12 G_2^{st}$, where $G^{st}_2 \approx
0.012 $ GeV$^4$ \cite{6} (see \cite{9} for a recent  gauge-string
duality treatment of $G^{st}_2$ yielding $G^{st}_2 \approx (0.01
\pm 0.002)$ GeV$^4$).

This $\Delta G_2$ taken as the change of free energy (pressure)
across the phase boundary will be our basic element in finding the
phase transition curve $T_c(\mu)$ below.

To this end we introduce in the next section  the NP EoS of qgp
derived recently in \cite{10}, and express $T_c(\mu)$ in terms of
$\Delta G_2$ and $L_{fund}(T_c)$.

Taking for the latter the lattice or analytic value, one obtains a
set of curves $T_c(\mu)$ for $n_f=0,2,3$ depending on the only
parameter $\Delta G_2$. These resulting  curves  and their end
points $T_c(0), \mu_c(0)$ are  discussed in conclusion.

2. In the NP approach to the  qgp in \cite{10} one introduces in
the first approximation the interaction of single quarks and
gluons with the vacuum, which is called the Single Line
Approximation (SLA), leaving pair and triple, etc... correlations
to the next steps. As a result one obtains in SLA the pressure
$P_q^{SLA}$ of quarks (and antiquarks) and $P_g^{SLA}$ of gluons
which are expressed through $L_{fund}$, namely \cite{10}.

 \be p_q\equiv{\frac{
P^{SLA}_{q}}{T^4}} = \frac{4N_cn_f}{{\pi}^2}
\sum^\infty_{n=1}\frac{ (-1)^{n+1}}{n^4} L^{n}_{fund}
\varphi^{(n)}_q cosh\frac{\mu n}{T}\label{4}\ee

\be p_{gl} \equiv \frac{P_{gl}^{SLA}}{T^4}
=\frac{2(N^2_c-1)}{\pi^2} \sum^\infty_{n=1} \frac{1}{n^4}
L_{adj}^n \label{5}\ee with \be \varphi^{(n)}_q (T)
=\frac{n^2m^2_q}{2T^2} K_2 \left(\frac{m_q n}{T}\right)\approx
1-\frac{1}{4} \left(\frac{n m_q}{T}\right)^2+...\label{6}\ee

In (\ref{4}), (\ref{5}) it was assumed that $T\la
\frac{1}{\lambda}\cong 1$ GeV, where $\lambda$ is the vacuum
correlation length, e.g. $D_1^{(E)} (x) \sim e^{-|x|/\lambda}$,
hence powers of $L^n_i$,  see \cite{10} for details.

With few percent accuracy one can replace the sum in (\ref{5}) by
the first term, $n=1$, and this form will be used below for
$p_{gl}$, while for $p_q$ this replacement is not valid for large
$\frac{\mu}{T}$, and one can use instead the form equivalent to
(\ref{4}), \be p_q=\frac{n_f}{\pi^2} \left[ \Phi_\nu \left(
\frac{\mu-\frac{V_1}{2}}{T}\right)+ \Phi_\nu \left(-
\frac{\mu+\frac{V_1}{2}}{T}\right)\right]\label{7}\ee where
$\nu=m_q/T$ and \be \Phi_\nu (a) =\int^\infty_0  \frac{z^4
dz}{\sqrt{z^2+\nu^2}}\frac{1}{(e^{\sqrt{z^2+\nu^2}-a}+1)}.\label{8}\ee

Eqs. (\ref{7}), (\ref{5}) define $p_q, p_{gl}$ for all $T, \mu$
and $m_q$, which is the current (pole) quark mass at the scale of
the order of $T$.

Using (\ref{4}-(\ref{8}) we can define the pressure $P_I$ in the
confined phase, and $P_{II}$ in the deconfined phase, taking into
account that vacuum energy density in two phases
$\varepsilon_{vac} $ and $\varepsilon_{vac}^{dec}$ respectively
contributes to the free energy, and hence $|\varepsilon_{vac}|,
|\varepsilon_{vac}^{dec}|$ to  the pressure. One  has \be P_{I}
=|\varepsilon_{vac}| + P_{hadron}, ~~ P_{II} =
|\varepsilon_{vac}^{dec}| + (p_{gl} + p_q) T^4.\label{9}\ee Here
$P_{hadron}$ is  the pressure of the hadron gas at $T\leq T_c$.
From $P_I(T_c) =P_{II}(T_c)$ one obtains $T_c$, neglecting
$P_{hadron}$ in the first approximation

\be T_c(\mu) =\left(\frac{(11 -\frac23 n_f)\Delta G_2}{32
(p_{gl}+p_q)}\right)^{1/4}.\label{10}\ee

In (\ref{10}) enter only two parameters; $\Delta G_2$ and
$L_{fund} (T_c) =\exp \left( -\frac{\kappa}{T_c}\right)$, $\kappa
\equiv\frac12 {V_1(\infty, T_c)}\cong \frac12 F^1_{Q\bar Q}(
\infty, T_c)$.

The latter can be found in 3 different ways: 1) from the direct
lattice measurements \cite{11} of $P^1_{Q\bar Q}\approx 0.5$ GeV;
2) from analytic calculation of $D_1^E$ in \cite{8}, which yields
$V_1(\infty, T<T_c)\approx \frac{6\alpha_s(M_0)\sigma}{M_0}\approx
0.5$ GeV with  $M_0\cong (0.8-1) $ GeV lowest gluelump mass
\cite{12}; 3) from lattice calculations of $D_1^E $ at  $T>T_c$,
\cite{7,13}, which according to (\ref{2}) yields  $V_1(\infty,
T_c) \approx 0.4 \div 0.6$ GeV. Therefore one can fix $V_1(\infty,
T_c)=0.50(5)$ GeV  ($\kappa=0.25$ GeV) and this value is
independent of $n_f$ \cite{11}. As a result $T_c(\mu)$ is a
function of only $\Delta G_2$ and  for each value of $\Delta G_2$
one finds a set of curves for $n_f =2,3,...$ We choose $\Delta
G_2\approx \frac12 G_2^{st}$ and in Fig.1  the curves computed
numerically  from (\ref{10}) for $n_f=2,3$ are shown for $\Delta
G_2\approx 0.00341 $ GeV$^4$ and zero quark pole masses.

\begin{figure}[htb]
\includegraphics[height=7cm]{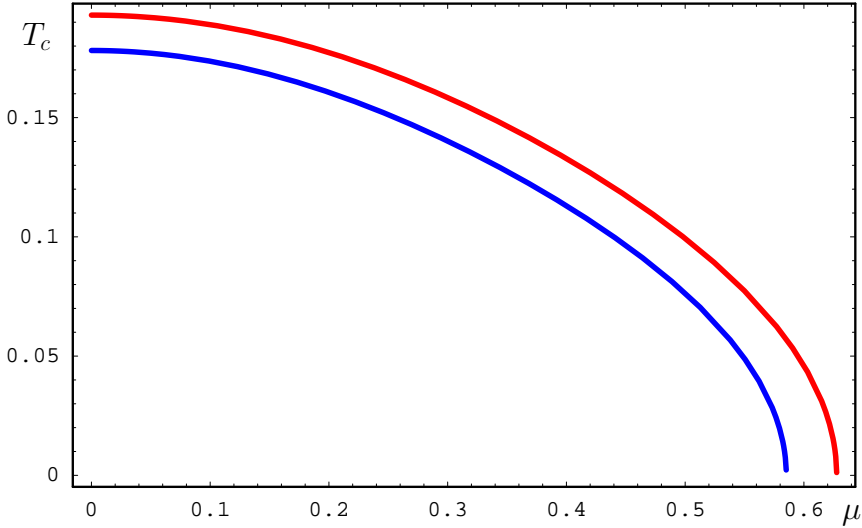}
\caption{The phase transition curve $T_c(\mu)$ from Eq.(\ref{10})
(in GeV) as function of quark chemical potential $\mu$ (in GeV)
for $n_f=2$ (upper curve) and $ n_f=3$ (lower curve) and $\Delta
G_2=0.0034$ GeV$^4$.}.
\end{figure}

The end points $T_c(0)$ and $\mu_c(0)$ can be found analytically
e.g. for $T_c(\mu)$ with $\sim 5\%$ accuracy  one has (expanding
(\ref{10}) in $\frac{p_{gl}}{p_q})$ \be T_c(\mu) = T_c(0)
\left(1-C\frac{9\mu^2}{T^2_c(0)}\right), ~~
C=0.0110(3),\label{11}\ee with \be T_c(0) \approx \frac12 T^{(0)}
\left( 1+\sqrt{1+\frac{\kappa}{T^{(0)}}}\right),~~ T^{(0)}=
\left(\frac{(11 -\frac23 n_f)\pi^2\Delta G_2}{384 n_f}
\right)^{1/4}.\label{12}\ee

For $\Delta G_2 = 0.00341$ GeV$^4$ one obtains $T_c(0) = (0.27;
0.19; 0.17)$ GeV for $ n_f =0,2,3$ respectively, which agrees well
with numerous lattice data, see \cite{14} for reviews. The value
of $C=0.011$ is inside the scattered set of  lattice values
\cite{14}.

Another end point, $\mu_c(0)$ can be found from the asymptotics of
(\ref{8}), $\Phi_0(a\to \infty) =\frac{a^4}{4} + \frac{\pi^2}{2}
a^2+...,$ which yields (for  $m_q=0)$ $\mu_c(0) =\kappa +
(48)^{1/4} T^{(0)}$ and for the same $\Delta G_2$ as above one
gets $\mu_c(0)=(0.63; 0.58)$ GeV for $ n_f=2,3$ One can check,
that the derivative in $T$, $\frac{d\mu_c(T)}{dT}$ vanishes at
$T=0$.

3. The phase curve  $T_c(\mu)$ in Fig.1 is in reasonable agreement
with lattice data at least for $\mu\la 0.25$ GeV, see \cite{15}
for review and references. Two important points are to be
discussed here: 1) order of transition and  possible critical
point 2) approximations and assumptions of the present work.

1). The vacuum transition of our approach is evidently  of the
first order at least in the leading (SLA) approximation used for
(\ref{10}), and does not contain any critical points. This is  in
agreement with lattice $n_f=0$ data, but the  lattice results for
$n_f=2,3$ depend on masses, discretization and are not fully
conclusive. The softening of transition for $n_f>0$ in our
approach is explained by the increasing role of $P_{hadron} (T)$
for $n_f>0$  near $T_c$, which suppresses the  specific heat and
makes the curve $P(T)$ more smooth. The chiral transition in our
approach is caused by the deconfinement, since  both $\lan \bar q
q\ran$ and $f_\pi$ are expressed via $D^E(x)$ \cite{16} and vanish
together   with it, in agreement  with lattice data, see e.g.
\cite{14}. The Polyakov loop is  a good   (approximate) order
parameter for $n_f =0(n_f>0)$ since at $T<T_c$ it is expressed via
$D^E(x)$ and vanishes (strongly decreases) (see Eq. (\ref{6}) of
\cite{5}).

2) In our derivation of (\ref{10})-(\ref{12}) it was assumed a)
that the only important part of qgp dynamics is the interaction
with the NP vacuum --SLA; b) it is   assumed that vacuum fields do
not  depend on $T, \mu$ in the  phase diagram, except  at the
phase boundary  where the shift $\Delta G_2$ occurs; in particular
neither $\Delta G_2$ nor $L_{fund} (T_c)$ depend on $\mu$. The
latter point is partly supported by lattice data \cite{17}. In
general this picture of rigid vacuum is based on the  notion of
the dilaton scale $m_d$ of vacuum fields, which can be associated
with  the $0^{++}$ glueball mass around 1.5 GeV and therefore for
all external parameters (like $\mu$ or $T$) much less than $m_d$
vacuum fields are fixed.

Another argument in favor of rigid vacuum  is that all dependence
on $\mu$ and $n_f$ does not appear in the lowest order of $1/N_c$
expansion, since it comes from  the quark loops. (Note that
nevertheless $T_c(0)$ differs strongly for $n_f=0$ and $n_f=2,3$;
even through $\Delta G_2$ was kept fixed, and this successful
prediction of $T_c(0)=0.27$ GeV and 0.19 GeV respectively can be
considered as another support of our picture).  Several things
were  not taken into account. Quark masses are included trivially
via  Eqs.(\ref{6},\ref{7}) and this can be checked $vs$ lattice
data. Phase transition near  $\mu_c(0)$ can be complicated due to
strong $qq$ and $qqq$ interaction, which is not taken into account
above and will be discussed elsewhere (see also \cite{10}).
therefore the possibility of color superconductivity is not
commented here.

The authors are grateful for useful discussions to members of ITEP
seminar. The financial support of RFBR grants 06-02-17012,  grant
for scientific schools NSh-843.2006.2 and the State Contract
02.445.11.7424 is gratefully acknowledged. This work was done with
financial support of the Federal Agency of Atomic Energy.

\end{document}